\documentclass[prl,aps,floatfix,showpacs,twocolumn,reprint,superscriptaddress]{revtex4-1}
\usepackage{amssymb,amsfonts,amsmath}
\usepackage{graphics}
\usepackage[dvips]{graphicx}
\usepackage{color}
\usepackage{subfigure}
\usepackage{mathtools}
\usepackage{multirow}
\usepackage{booktabs}
\usepackage{bm}
\usepackage[linktocpage,colorlinks=true,linkcolor=blue,citecolor=blue]{hyperref}

\begin{document}


\title{Non-renewal statistics in the catalytic activity of enzyme molecules at mesoscopic concentrations}

\author{Soma Saha}
\affiliation{Department of Chemistry, Indian Institute of Technology, Madras, Chennai 600036, India}
\author{Somdeb Ghose}
\affiliation{The Institute of Mathematical Sciences, CIT Campus, Tharamani, Chennai-600113, India}
\author{R. Adhikari}
\affiliation{The Institute of Mathematical Sciences, CIT Campus, Tharamani, Chennai-600113, India}
\author{Arti Dua}
\affiliation{Department of Chemistry, Indian Institute of Technology, Madras, Chennai 600036, India}

\date{\today}

\begin{abstract}
Recent fluorescence spectroscopy measurements of single-enzyme kinetics have shown that enzymatic turnovers form a renewal stochastic process in which the inverse of the mean waiting time between turnovers follows the Michaelis-Menten equation. Under typical physiological conditions, however, tens to thousands of enzymes react in catalyzing thousands to millions of substrates. We study enzyme kinetics at these physiologically relevant conditions through a master equation including stochasticity and molecular discreteness. From the exact solution of the master equation we find that the waiting times are neither independent nor are they identically distributed, implying that enzymatic turnovers form a non-renewal stochastic process. The inverse of the mean waiting time shows strong departures from the Michaelis-Menten equation. The waiting times between consecutive turnovers are anti-correlated, where short intervals are more likely to be followed by long intervals and vice versa. Correlations persist beyond consecutive turnovers indicating that multi-scale fluctuations govern enzyme kinetics. 
\end{abstract}

\maketitle

Biological processes rely crucially on the catalytic activity of enzymes. In 1913, following the work of Wurtz and several others \cite{wurtz1880,*osullivan1890,*brown1902,*henri1902}, Michaelis and Menten proposed \cite{michaelis1913} a reaction mechanism for catalysis where enzyme $E$ binds reversibly with substrate $S$ to form an enzyme-substrate complex $ES$ which then dissociates irreversibly to form product $P$, while regenerating the enzyme: $E + S \xrightleftharpoons[k_{-1}]{k_1} ES \xrightarrow{k_2} E + P.$
%
%
For thermodynamically large numbers of reactants, deterministic mass action kinetics provides the temporal variation of the concentrations of enzyme, complex and product. The rate of product formation is given by the classic Michaelis-Menten (MM) equation, provided suitable adiabaticity conditions are satisfied \cite{segel1988,*segel1989}.

However, enzyme and substrate concentrations in biochemical catalysis are not thermodynamically large. \emph{In vivo} enzyme concentrations vary from nanomolar to micromolar, while the substrates are typically between a ten and ten thousand times more numerous \cite{albe1990}. An important exception is in glycolysis where substrate concentrations exceed those of enzymes \cite{albe1990}. \emph{In vitro} enzyme concentrations vary from picomolar to nanomolar and substrates are typically a million times more numerous \cite{tinoco2002,*schnell2003}. At these low concentrations, the inherent stochasticity of a single chemical reaction and the discrete change in the number of reactant molecules combine to generate spontaneous, intrinsic fluctuations known as molecular noise \cite{mccullagh2009}. The temporal variation of catalysis, then, is also influenced by molecular noise and is a stochastic process in time. Recent advances in single molecule spectroscopy have been able to unravel some features of this stochastic process for catalysis involving a single enzyme and numerous substrates \cite{kou2005,english2006}. A striking feature is that the enzymatic turnovers generate  a renewal point process where the waiting time $\tau$ between product formation events is independently and identically distributed. Remarkably, the inverse of the mean waiting time $\langle \tau \rangle ^{-1}$ obeys the MM equation which, in this interpretation, is valid not only for thermodynamically large systems, but also at the single-enzyme level.

In this Letter, we study the stochastic process of enzymatic turnovers at concentrations between the extremes of the thermodynamically large and single-enzyme regimes. In the thermodynamic limit the process reduces to deterministic evolution  governed by mass action kinetics, while in the single-enzyme limit it reduces to a renewal process. Our key findings are that for mesoscopic numbers of enzymes, the turnover process is of the non-renewal type with waiting times that are neither independent, nor identically distributed.  We calculate the waiting time distributions and show that their inverse first moments do not obey the MM equation. Consecutive waiting times are anti-correlated, with short intervals more likely to be followed by long intervals and vice-versa. The correlations persist beyond consecutive turnovers and, depending on the number of enzymes, can become substantially long-ranged. Together, these results imply that the enzymatic turnovers at the mesoscale cannot be described by mean production rates (as in the thermodynamic limit) or mean waiting times (as in the single-enzyme limit), but must be described by statistical measures which capture fluctuations over multiple time scales. 

\begin{figure}[]
\centering
 \includegraphics[width=0.35\textwidth]{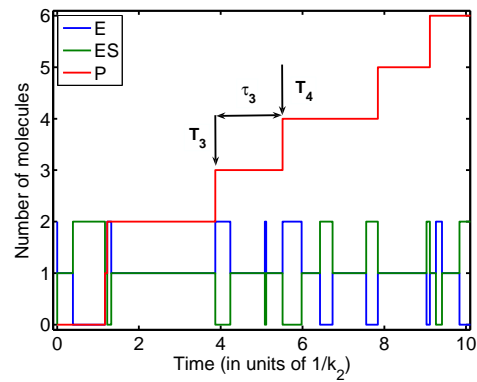}
 \caption{(Color online) A trajectory  of Eq.\ (\ref{eq:CME}) for $N=2$ enzymes. The $p$-th product is generated at time $T_p$. The waiting time between the $p$-th and  $(p+1)$-th product is $\tau_p = T_{p+1} - T_p$.}  
\label{fig:timetrace}
\end{figure}

\emph{Model.}--- We begin by describing the catalytic process through $P(n_E, n_{ES}, n_P, t)$, the joint probability that there are  $n_E$ enzymes, $n_{ES}$ enzyme-substrate complexes and $n_P$ products at any time $t$, starting initially with $N$ enzymes, $S$ substrates and no complexes or products. Assuming that the system is well-mixed, the probability is taken to obey the Markovian chemical master equation
\begin{eqnarray} \label{eq:CME}
\dot P(n_E, n_{ES}, n_P, t) = \nonumber \\
 k_a (n_E+1)\, P(n_E+1, n_{ES}-1, n_P, t) \nonumber\\
+ k_{-1} (n_{ES}+1)\, P(n_E-1, n_{ES}+1, n_P, t)\nonumber\\
+ k_{2} (n_{ES}+1)\, P(n_E-1, n_{ES}+1, n_P-1, t) \nonumber \\
-\Big[k_a n_E + (k_{-1} + k_{2} ) n_{ES}\Big]\,P(n_E, n_{ES}, n_{P}, t)
\end{eqnarray}
with the transition rates chosen to describe the MM kinetics of the enzyme catalysis reaction system given earlier. Since substrates are more numerous than enzymes, the bimolecular second-order complexation step $E + S \xrightarrow{k_1} ES$ is replaced by a pseudo-first order complexation step with an effective rate constant $k_a = k_1 S$. The master equation generates stochastic trajectories of the kind shown in Fig.\ (\ref{fig:timetrace}). Since enzymes are either converted to the enzyme-substrate complex or are regenerated from it, physical trajectories obey the constraint $n_E + n_{ES} = N$ at all times. The probability distribution of these trajectories can then be written as  $P(N -n_{ES}, n_{ES},n_P,t|N,S)$ which we abbreviate to $P(n_{ES},n_P,t|N)$. This simplifies the solution as there are two,  and not three, independent variables. 

\emph{Exact solution}.--- We use the generating function method to obtain an exact solution of the master equation. A related solution with $n_E$ and $n_P$ as independent variables is given in \cite{heyde1969}. Defining the generating function as \vspace{-0.5em}
\begin{align} \label{eq:gf1}
 G(s_1, s_2, t ) = \sum _{n_{ES}}\sum_{n_P} s_{1} ^{n_{ES}}  s_{2} ^{n_P}\,  P(n_{ES}, n_P, t | N)
\end{align}
we find from the master equation its equation of motion \cite{vanKampen2007}, \vspace{-0.5em}
\begin{eqnarray}\label{eq:gf2}
\lefteqn{ \partial_t G(s_{1}, s_{2}, t)= k_{a} N (s_{1}-1) G(s_{1}, s_{2}, t) \, \, +}  \nonumber \\
&& \left[(1 - s_{1}) (k_{b} + k_{a} s_{1}) -k_2 (1 - s_2)\right] \partial_{s_{1}} G(s_1, s_{2}, t),
\end{eqnarray}
where $k_{-1} + k_{2} = k_{b}$. This partial differential equation in $s_1, s_2$ and $t$ can be solved by the method of characteristics which, after a lengthy calculation, yields
\begin{eqnarray} \label{eq:gf3}
 \lefteqn{ G(s_{1}, s_2, t) = \frac{1}{2^{N}} \bigg\{ e^{-(B - A')t} + e^{-(B + A')t}\bigg.} \nonumber \\
 && + \frac{k_{b}-k_{a} (1 - 2 s_1)}{2 A'} \Big[e^{-(B - A')t}-e^{-(B + A')t}\Big]\bigg\}^{N},
\end{eqnarray}
where $ A' = \frac{1}{2}\sqrt{(k_{a} + k_{b})^{2} - 4k_{a} k_2 (1 - s_2)}$ and $B = \frac{1}{2}(k_a+k_b)$. $P(n_{ES},n_P,t|N)$, obtained from the coefficient of the Taylor expansion of the generating function in the $s_1$ and $s_2$ variables, is an exact solution to the master equation. To support and complement this exact analytical solution, we generate exact numerical trajectories of Eq.\ (\ref{eq:CME}) using the Doob-Gillespie algorithm \cite{doob1945,*gillespie1976,*gillespie1977}. For the numerical simulations we non-dimensionalize time in units of $k_2$ and choose rate constants as $k_a = k_2$ and $k_{-1} = \frac{1}{2} k_2$. We generate ensembles of typically $10^6$ trajectories to obtain the probability distributions of interest. One such trajectory is shown in Fig.\ (\ref{fig:timetrace}).

\emph{Turnover statistics}.--- The trajectories in  Fig.\ (\ref{fig:timetrace}) can be described in either of two ways : we can count the number of enzymatic turnover events $n_P$  that have occurred in duration of time $t$, or, we can specify the time $T_p$ at which the $p$-th turnover occurs. The former, called the counting process description \cite{daley2003}, has been used in most previous studies as it follows directly from the solution $P(n_{ES},n_P,t|N)$ of the master equation. The latter, called the point process description \cite{bhabha1950,*ramakrishnan1950}, has not (to the best of our knowledge) been studied before for multiple enzymes. This is the focus of our work. 

We define turnover times as $T_p = \mathrm{inf}\{t > 0 : n_P(t) \geq p\}$ for $p=1, 2, \ldots $, which implies that $T_p \leq t$ if and only if $n_P(t) \geq p$. This provides the connection between the counting and point processes and relates the cumulative distribution of $T_p$ to that of $n_P$ by $P(T_p \leq t)=P(n_p \geq p, t)$ \cite{daley2003}. Waiting times are defined from the turnover times by $\tau_p = T_{p} - T_{p-1}$ with the convention that $T_0 =0$. The point process is fully specified by the joint probability distributions $w$ of either the $T_p$ or the $\tau_p$ \cite{bhabha1950,*ramakrishnan1950,daley2003}.  Here we focus on the first-order distributions of the time to the $p$-th turnover $w(T_p)$ and the interval between the $p$-th and $(p+1)$-th turnovers $w(\tau_p)$. We use second-order distributions $w(\tau_{p},\tau_{p+q})$ to study correlations between the $p$-th and $(p+q)$-th turnovers.
\begin{figure}[tbp]
\centering
  \includegraphics[width=.35\textwidth]{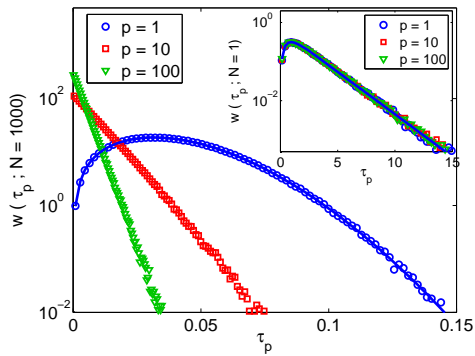}
  \caption{(Color online) Waiting time distributions for $N=1000$ and $N=1$ enzymes. The waitings times are identically distributed for a single enzyme (inset), but vary with the turnover number $p$ for multiple enzymes. Solid lines are analytical results obtained from Eq.\ (\ref{eq:FPT:N}) while the symbols are simulation data.}
\label{fig:int}
\end{figure}
%

\emph{First-order distributions}.--- We derive exact expressions for $w(T_p)$ from the master equation solution using the connection between the counting and point processes. It follows that $ P(T_p \leq t) = P(n_P \geq p, t) = 1 - P(n_P < p, t) = 1 - \sum_{n_P=0}^{p-1}P(n_P, t)$. Since $P(n_P,t) = \sum_{n_{ES}}P(n_{ES}, n_{P}, t|N)$, it follows by differentiation that
%
%
\begin{equation}\label{eq:FPTeq}
  w(T_p) = -  \sum_{n_{p} = 0}^{p-1}\left. \sum_{n_{ES}}\partial_t P(n_{ES}, n_P, t | N)\right|_{t=T_p}.
\end{equation}
For fixed $n_p$, the terms in the inner summation are related to the time derivative of $G(s_1, s_2, t)$ evaluated at $s_1=1$ through Eq.\ (\ref{eq:gf1}). The equation of motion, Eq.\ (\ref{eq:gf2}), is then used to eliminate the time derivative of $G(s_1, s_2, t)$ in favour of its $s_1$ derivative. From this, each term of the outer summation is obtained by taking $n_P$ derivatives with respect to $s_2$, setting $s_2 = 0$ and then summing over $n_P$ to give \cite{SI}
%
 \begin{align} \label{eq:FPTdist}
 w(T_p) \! = \! \frac{k_2}{(p-1)!} \! \left[ \partial_{s_2}^{\,p-1} \partial_{s_1} G(s_{1}, s_2 , T_p| N) \right]_{s_1 = 1, s_2 = 0}.
 \end{align}
Using Eq.\ (\ref{eq:gf3}), $w(T_1) = k_2\left[\partial_{s_1}G(s_1,0,T_1|N)\right]_{s_1=1}$ is obtained as
\begin{eqnarray} \label{eq:FPT:N}
\lefteqn{ w(T_1) \! = \! \frac{k_2 k_a N}{(2 A)^N} \!\! \left[ e^{( A - B ) T_1} - e^{-( A + B ) T_1}\right] } \nonumber \\
&& \times \left[ ( A + B) e^{(A - B) T_1} + (A - B) e^{- ( B + A ) T_1}\right]^{N-1} 
\end{eqnarray}
and this is identical to $w(\tau_1)$. Here $A = \frac{1}{2}\sqrt{(k_{a} + k_{b})^{2}-4k_{a} k_2}$ and $B = \frac{1}{2}(k_a+k_b)$. Since higher waiting times are differences of consecutive turnover times, the joint probability of consecutive turnovers is needed to calculate the $w(\tau_p)$ for $p > 1$. This requires an involved calculation which we bypass by directly computing waiting time distributions from simulation trajectories. 

In Fig.\ (\ref{fig:int}) we compare $w(\tau_p)$ for $p=1, 10, 100$ for a reaction with $N=1000$ enzymes. The $\tau_p$  are not identically distributed. There is excellent agreement between the numerical result and the analytical calculation  $\tau_1$,  Eq.\ (\ref{eq:FPT:N}),  plotted as a solid line. In the inset, we plot  $w(\tau_p)$, but for a single enzyme. The $\tau_P$ are identically distributed, and agree with the analytical expression in Eq.\ (\ref{eq:FPT:N}) with $N=1$. This clearly establishes the non-renewal nature of the turnover process when more than one enzyme participates in catalysis. 

Surprisingly, starting with a Markovian master equation where waiting times between transitions are exponentially distributed, we obtain a waiting time between turnovers that is multi-exponential. For this, it is crucial to have more than one enzyme in the system. Then, as Eq.\ (\ref{eq:FPTeq}) shows,  multiple internal states for the enzyme-substrate complex have to be summed over, and the resulting point process for the products alone is no longer Markovian. The multi-exponentiality of the waiting times is, therefore, consistent with the non-Markovian nature of the turnovers. For a single enzyme, though, with only one internal enzyme-substrate state, there is no multi-exponentiality, but only a mono-exponential rise and fall. This is in agreement with earlier experimental \cite{lu1998,english2006}, numerical \cite{kou2005} and analytical \cite{lu1998} results. 
\begin{figure} 
 \centering 
  \includegraphics[width=.35\textwidth]{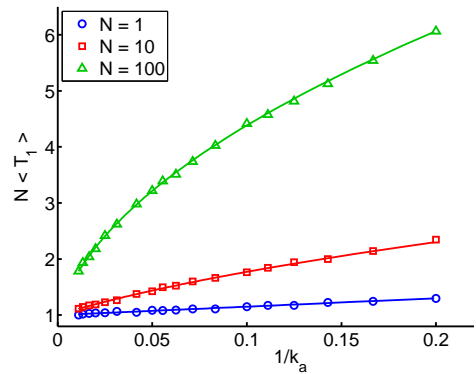} 
  \caption{\label{fig:LB}(Color online) First moment of $T_1$ plotted, in Lineweaver-Burk fashion, against the inverse rate constant $k_a$. Solid lines are first moments of  Eq.\ (\ref{eq:FPT:N}) while symbols are simulation data. With more than one enzyme, the MM equation (circles) is obeyed only in the limit of infinite substrate concentration or equivalently for $1/k_a \rightarrow 0$.}
\end{figure}

\emph{Moments of first-order distributions}.--- For a single enzyme, it follows from Eq.\ (\ref{eq:FPT:N}) that $\langle T_1 \rangle = \int_0^{\infty} dT_1~T_1 w(T_1) = (S + K_{M})/k_2 S$. The inverse of $\langle T_1 \rangle$ then obeys the MM equation which has lead Xie and coworkers to extend the validity of the  MM equation to the single-enzyme level \cite{lu1998,kou2005,english2006}. However, for multiple enzymes, we find that the first moment no longer obeys the Michaelis-Menten equation as can be seen in in Fig.\ (\ref{fig:LB}), where we plot $N \langle T_1 \rangle$ against $1/k_a$ in Lineweaver-Burk fashion. Thus, a turnover time interpretation of the MM equation is no longer valid for multiple enzymes.

If $N$ independent single-enzyme MM renewal process trajectories were to be pooled, there would be an $N$-fold decrease in the mean turnover times. Fig.\ (\ref{fig:LB}) shows that $\langle T_1\rangle$ is larger than the MM estimate, indicating a slowing down of the kinetics due to cooperativity. The mean turnover time converges to the MM estimate only in the limit of infinite substrate concentration or equivalently for $1/k_a \rightarrow 0$. Means $\langle T_p \rangle$ show similar behavior.

The non-linearity in $\langle T_1 \rangle$ is arises from the multi-exponentiality of $w(T_1)$. For large N, there is no closed form analytical expression for the mean turnover time. However, in the limit of $(k_a + k_b)^2 \gg 4 k_a k_2$, which amounts to the steady-state approximation in the deterministic kinetics, the expression for the mean turnover time is given by
\begin{eqnarray} \label{eq:mwtss}
&& \langle T_1 \rangle \approx \frac{1}{ N \delta} \left[ 1 - \frac{(N-1) N^2 \delta^6}{k_a k_2 (k_a k_2 + (N-1)\delta^2)^2} \,\, - \right. \nonumber\\
& &  \left.  \frac{N^2 \delta^4}{(k_a k_2 + (N-1)\delta^2 )^2}  \right. \!\!+\!\! \left. \frac{(N-1) N^2 \delta^6}{k_a k_2(2 k_a k_2 + (N-2)\delta^2)^2} \right] 
\end{eqnarray}
This is obtained by using $A - B \approx -\delta$, $A + B \approx 2B$ and $\delta = {k_a k_2}/{(k_a + k_b)}$ in Eq.\ (\ref{eq:FPT:N}), Taylor expanding to first order in $\left( \delta/2B \right) \exp{[-(2B - \delta)T_1]}$ and then computing the first moment of the approximated distribution \cite{SI}. The negative sign of the leading order correction term explains the curvature of the plot in Fig.\ (\ref{fig:LB}).  
\begin{figure}
\centering
 \includegraphics[width=.40\textwidth]{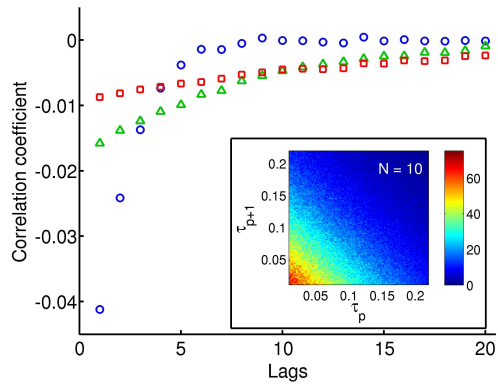} 
 \caption{The Pearson correlation coefficient of $\tau_p$ and $\tau_{p+q}$ as a function of lag $q$ for $N=10$ (circles), $N=50$ (triangles) and $N=100$ (squares). The inset shows the joint distribution $w(\tau_p, \tau_{p+1})$ for $N=10$.}
\label{fig:CPD}
\end{figure} 

\emph{Second-order distributions and memory}.--- We compute the joint distributions $w(\tau_{p},\tau_{p+q})$ of $\tau_p$ and $\tau_{p+q}$ and their Pearson correlation coefficient from numerical trajectories. In Fig.\ (\ref{fig:CPD}) we plot the correlation coefficient against lag $q$, showing the joint distribution of consecutive intervals in the inset. The waiting times are anti-correlated, where a short first interval is more likely to be followed by a long second interval and vice-versa. This memory effect shows a systematic variation with enzyme number, being strong and short-lived for fewer enzymes but weak and long-lived for more enzymes. With long-lived memory, fluctuation statistics will vary with the size of the temporal window, and multiple measures will be required to characterize the turnover process. In future work, we plan to explore this systematically, by studying higher-order joint distributions. The overall effect of the anti-correlations is to reduce the variance in the product turnovers when compared with a Poisson process. This may be biologically relevant to ensure a uniform rate of turnover in the steady state. 

\emph{Conclusion}.---The non-renewal properties of enzymatic turnovers presented here can be verified by fluorescence experiments with well-mixed reactants. Fluctuations of intermediate states which must be summed over lead to multi-exponential waiting time distributions for the product and to the correlations between waiting times. These non-renewal aspects should appear in other models of catalysis which involve several types of enzyme-substrate intermediates. Fluctuations of intermediate states can also provide a model for dynamic disorder, which has previously been modelled by fluctuating reaction rates. For second-order kinetics with substrate fluctuations \cite{bartholomay1962}, we numerically compute low order waiting time distributions and find negligible differences with our results. This justifies our use of pseudo-first order kinetics, which remains a reliable approximation at early times even when substrates fluctuate. In conclusion, the main implication of our work is that enzyme kinetics must be approached as a non-renewal stochastic process in time with fluctuations at multiple time scales. 

Financial support from the University Grants Commission (UGC), Government of India (SS) and PRISM, Department of Atomic Energy, Government of India (SG and RA) is gratefully acknowledged. The authors thank Indrani Bose, Binny Cherayil, Daan Frenkel, Wei Min, and Pieter Rein ten Wolde for helpful comments.

\bibliography{enzyme}

\begin{thebibliography}{24}%
\makeatletter
\providecommand \@ifxundefined [1]{%
 \@ifx{#1\undefined}
}%
\providecommand \@ifnum [1]{%
 \ifnum #1\expandafter \@firstoftwo
 \else \expandafter \@secondoftwo
 \fi
}%
\providecommand \@ifx [1]{%
 \ifx #1\expandafter \@firstoftwo
 \else \expandafter \@secondoftwo
 \fi
}%
\providecommand \natexlab [1]{#1}%
\providecommand \enquote  [1]{``#1''}%
\providecommand \bibnamefont  [1]{#1}%
\providecommand \bibfnamefont [1]{#1}%
\providecommand \citenamefont [1]{#1}%
\providecommand \href@noop [0]{\@secondoftwo}%
\providecommand \href [0]{\begingroup \@sanitize@url \@href}%
\providecommand \@href[1]{\@@startlink{#1}\@@href}%
\providecommand \@@href[1]{\endgroup#1\@@endlink}%
\providecommand \@sanitize@url [0]{\catcode `\\12\catcode `\$12\catcode
  `\&12\catcode `\#12\catcode `\^12\catcode `\_12\catcode `\%12\relax}%
\providecommand \@@startlink[1]{}%
\providecommand \@@endlink[0]{}%
\providecommand \url  [0]{\begingroup\@sanitize@url \@url }%
\providecommand \@url [1]{\endgroup\@href {#1}{\urlprefix }}%
\providecommand \urlprefix  [0]{URL }%
\providecommand \Eprint [0]{\href }%
\providecommand \doibase [0]{http://dx.doi.org/}%
\providecommand \selectlanguage [0]{\@gobble}%
\providecommand \bibinfo  [0]{\@secondoftwo}%
\providecommand \bibfield  [0]{\@secondoftwo}%
\providecommand \translation [1]{[#1]}%
\providecommand \BibitemOpen [0]{}%
\providecommand \bibitemStop [0]{}%
\providecommand \bibitemNoStop [0]{.\EOS\space}%
\providecommand \EOS [0]{\spacefactor3000\relax}%
\providecommand \BibitemShut  [1]{\csname bibitem#1\endcsname}%
\let\auto@bib@innerbib\@empty
\bibitem [{\citenamefont {Wurtz}(1880)}]{wurtz1880}%
  \BibitemOpen
  \bibfield  {author} {\bibinfo {author} {\bibfnamefont {C.~A.}\ \bibnamefont
  {Wurtz}},\ }\href@noop {} {\bibfield  {journal} {\bibinfo  {journal} {Compt.
  Rend. Hebd. Acad. Sci. Paris}\ }\textbf {\bibinfo {volume} {91}},\ \bibinfo
  {pages} {787} (\bibinfo {year} {1880})}\BibitemShut {NoStop}%
\bibitem [{\citenamefont {O'Sullivan}\ and\ \citenamefont
  {Tompson}(1890)}]{osullivan1890}%
  \BibitemOpen
  \bibfield  {author} {\bibinfo {author} {\bibfnamefont {C.}~\bibnamefont
  {O'Sullivan}}\ and\ \bibinfo {author} {\bibfnamefont {F.}~\bibnamefont
  {Tompson}},\ }\href@noop {} {\bibfield  {journal} {\bibinfo  {journal} {J.
  Chem. Soc., Trans.}\ }\textbf {\bibinfo {volume} {57}},\ \bibinfo {pages}
  {834} (\bibinfo {year} {1890})}\BibitemShut {NoStop}%
\bibitem [{\citenamefont {Brown}(1902)}]{brown1902}%
  \BibitemOpen
  \bibfield  {author} {\bibinfo {author} {\bibfnamefont {A.}~\bibnamefont
  {Brown}},\ }\href@noop {} {\bibfield  {journal} {\bibinfo  {journal} {J.
  Chem. Soc.}\ }\textbf {\bibinfo {volume} {81}},\ \bibinfo {pages} {373}
  (\bibinfo {year} {1902})}\BibitemShut {NoStop}%
\bibitem [{\citenamefont {Henri}(1902)}]{henri1902}%
  \BibitemOpen
  \bibfield  {author} {\bibinfo {author} {\bibfnamefont {V.}~\bibnamefont
  {Henri}},\ }\href@noop {} {\bibfield  {journal} {\bibinfo  {journal} {Compt.
  Rend. Hebd. Acad. Sci. Paris}\ }\textbf {\bibinfo {volume} {135}},\ \bibinfo
  {pages} {916} (\bibinfo {year} {1902})}\BibitemShut {NoStop}%
\bibitem [{\citenamefont {Michaelis}\ and\ \citenamefont
  {Menten}(1913)}]{michaelis1913}%
  \BibitemOpen
  \bibfield  {author} {\bibinfo {author} {\bibfnamefont {L.}~\bibnamefont
  {Michaelis}}\ and\ \bibinfo {author} {\bibfnamefont {M.~L.}\ \bibnamefont
  {Menten}},\ }\href {http://www.citeulike.org/group/2018/article/1550034}
  {\bibfield  {journal} {\bibinfo  {journal} {Biochem. Z}\ }\textbf {\bibinfo
  {volume} {49}},\ \bibinfo {pages} {333} (\bibinfo {year} {1913})}\BibitemShut
  {NoStop}%
\bibitem [{\citenamefont {Segel}(1988)}]{segel1988}%
  \BibitemOpen
  \bibfield  {author} {\bibinfo {author} {\bibfnamefont {L.}~\bibnamefont
  {Segel}},\ }\href {\doibase 10.1007/BF02460092} {\bibfield  {journal}
  {\bibinfo  {journal} {B. Math. Biol.}\ }\textbf {\bibinfo {volume} {50}},\
  \bibinfo {pages} {579} (\bibinfo {year} {1988})}\BibitemShut {NoStop}%
\bibitem [{\citenamefont {Segel}\ and\ \citenamefont
  {Slemrod}(1989)}]{segel1989}%
  \BibitemOpen
  \bibfield  {author} {\bibinfo {author} {\bibfnamefont {L.}~\bibnamefont
  {Segel}}\ and\ \bibinfo {author} {\bibfnamefont {M.}~\bibnamefont
  {Slemrod}},\ }\href {\doibase 10.1137/1031091} {\bibfield  {journal}
  {\bibinfo  {journal} {SIAM Rev.}\ }\textbf {\bibinfo {volume} {31}},\
  \bibinfo {pages} {446} (\bibinfo {year} {1989})}\BibitemShut {NoStop}%
\bibitem [{\citenamefont {Albe}\ \emph {et~al.}(1990)\citenamefont {Albe},
  \citenamefont {Butler},\ and\ \citenamefont {Wright}}]{albe1990}%
  \BibitemOpen
  \bibfield  {author} {\bibinfo {author} {\bibfnamefont {K.~R.}\ \bibnamefont
  {Albe}}, \bibinfo {author} {\bibfnamefont {M.~H.}\ \bibnamefont {Butler}}, \
  and\ \bibinfo {author} {\bibfnamefont {B.~E.}\ \bibnamefont {Wright}},\
  }\href {\doibase 10.1016/S0022-5193(05)80266-8} {\bibfield  {journal}
  {\bibinfo  {journal} {J. Theor. Biol.}\ }\textbf {\bibinfo {volume} {143}},\
  \bibinfo {pages} {163} (\bibinfo {year} {1990})}\BibitemShut {NoStop}%
\bibitem [{\citenamefont {Tinoco}\ \emph {et~al.}(2002)\citenamefont {Tinoco},
  \citenamefont {Sauer}, \citenamefont {Wang},\ and\ \citenamefont
  {Puglisi}}]{tinoco2002}%
  \BibitemOpen
  \bibfield  {author} {\bibinfo {author} {\bibfnamefont {I.}~\bibnamefont
  {Tinoco}}, \bibinfo {author} {\bibfnamefont {K.}~\bibnamefont {Sauer}},
  \bibinfo {author} {\bibfnamefont {J.~C.}\ \bibnamefont {Wang}}, \ and\
  \bibinfo {author} {\bibfnamefont {J.~D.}\ \bibnamefont {Puglisi}},\
  }\href@noop {} {\emph {\bibinfo {title} {Physical Chemistry: Principles and
  Applications in Biological Sciences}}},\ \bibinfo {edition} {4th}\ ed.\
  (\bibinfo  {publisher} {Prentice Hall},\ \bibinfo {address} {New Jersey},\
  \bibinfo {year} {2002})\BibitemShut {NoStop}%
\bibitem [{\citenamefont {Schnell}\ and\ \citenamefont
  {Maini}(2003)}]{schnell2003}%
  \BibitemOpen
  \bibfield  {author} {\bibinfo {author} {\bibfnamefont {S.}~\bibnamefont
  {Schnell}}\ and\ \bibinfo {author} {\bibfnamefont {P.~K.}\ \bibnamefont
  {Maini}},\ }\href
  {http://www.informatics.indiana.edu/schnell/papers/ctb8_169.pdf} {\bibfield
  {journal} {\bibinfo  {journal} {Comments Theor. Biol.}\ }\textbf {\bibinfo
  {volume} {8}},\ \bibinfo {pages} {169} (\bibinfo {year} {2003})}\BibitemShut
  {NoStop}%
\bibitem [{\citenamefont {McCullagh}\ \emph {et~al.}(2009)\citenamefont
  {McCullagh}, \citenamefont {Farlow}, \citenamefont {Fuller}, \citenamefont
  {Girard}, \citenamefont {Lipinski-Kruszka}, \citenamefont {Lu}, \citenamefont
  {Noriega}, \citenamefont {Rollins}, \citenamefont {Spitzer}, \citenamefont
  {Todhunter},\ and\ \citenamefont {El-Samad}}]{mccullagh2009}%
  \BibitemOpen
  \bibfield  {author} {\bibinfo {author} {\bibfnamefont {E.}~\bibnamefont
  {McCullagh}}, \bibinfo {author} {\bibfnamefont {J.}~\bibnamefont {Farlow}},
  \bibinfo {author} {\bibfnamefont {C.}~\bibnamefont {Fuller}}, \bibinfo
  {author} {\bibfnamefont {J.}~\bibnamefont {Girard}}, \bibinfo {author}
  {\bibfnamefont {J.}~\bibnamefont {Lipinski-Kruszka}}, \bibinfo {author}
  {\bibfnamefont {D.}~\bibnamefont {Lu}}, \bibinfo {author} {\bibfnamefont
  {T.}~\bibnamefont {Noriega}}, \bibinfo {author} {\bibfnamefont
  {G.}~\bibnamefont {Rollins}}, \bibinfo {author} {\bibfnamefont
  {R.}~\bibnamefont {Spitzer}}, \bibinfo {author} {\bibfnamefont
  {M.}~\bibnamefont {Todhunter}}, \ and\ \bibinfo {author} {\bibfnamefont
  {H.}~\bibnamefont {El-Samad}},\ }\href {\doibase 10.1038/nchembio.222}
  {\bibfield  {journal} {\bibinfo  {journal} {Nat. Chem. Biol.}\ }\textbf
  {\bibinfo {volume} {5}},\ \bibinfo {pages} {699} (\bibinfo {year}
  {2009})}\BibitemShut {NoStop}%
\bibitem [{\citenamefont {Kou}\ \emph {et~al.}(2005)\citenamefont {Kou},
  \citenamefont {Cherayil}, \citenamefont {Min}, \citenamefont {English},\ and\
  \citenamefont {Xie}}]{kou2005}%
  \BibitemOpen
  \bibfield  {author} {\bibinfo {author} {\bibfnamefont {S.~C.}\ \bibnamefont
  {Kou}}, \bibinfo {author} {\bibfnamefont {B.~J.}\ \bibnamefont {Cherayil}},
  \bibinfo {author} {\bibfnamefont {W.}~\bibnamefont {Min}}, \bibinfo {author}
  {\bibfnamefont {B.~P.}\ \bibnamefont {English}}, \ and\ \bibinfo {author}
  {\bibfnamefont {X.~S.}\ \bibnamefont {Xie}},\ }\href {\doibase
  10.1021/jp051490q} {\bibfield  {journal} {\bibinfo  {journal} {J. Phys. Chem.
  B}\ }\textbf {\bibinfo {volume} {109}},\ \bibinfo {pages} {19068} (\bibinfo
  {year} {2005})}\BibitemShut {NoStop}%
\bibitem [{\citenamefont {English}\ \emph {et~al.}(2006)\citenamefont
  {English}, \citenamefont {Min}, \citenamefont {{van Oijen}}, \citenamefont
  {Lee}, \citenamefont {Luo}, \citenamefont {Sun}, \citenamefont {Cherayil},
  \citenamefont {Kou},\ and\ \citenamefont {Xie}}]{english2006}%
  \BibitemOpen
  \bibfield  {author} {\bibinfo {author} {\bibfnamefont {B.~P.}\ \bibnamefont
  {English}}, \bibinfo {author} {\bibfnamefont {W.}~\bibnamefont {Min}},
  \bibinfo {author} {\bibfnamefont {A.~M.}\ \bibnamefont {{van Oijen}}},
  \bibinfo {author} {\bibfnamefont {K.~T.}\ \bibnamefont {Lee}}, \bibinfo
  {author} {\bibfnamefont {G.}~\bibnamefont {Luo}}, \bibinfo {author}
  {\bibfnamefont {H.}~\bibnamefont {Sun}}, \bibinfo {author} {\bibfnamefont
  {B.~J.}\ \bibnamefont {Cherayil}}, \bibinfo {author} {\bibfnamefont {S.~C.}\
  \bibnamefont {Kou}}, \ and\ \bibinfo {author} {\bibfnamefont {X.~S.}\
  \bibnamefont {Xie}},\ }\href {\doibase 10.1038/nchembio759} {\bibfield
  {journal} {\bibinfo  {journal} {Nat. Chem. Biol.}\ }\textbf {\bibinfo
  {volume} {2}},\ \bibinfo {pages} {87} (\bibinfo {year} {2006})}\BibitemShut
  {NoStop}%
\bibitem [{\citenamefont {Heyde}\ and\ \citenamefont
  {Heyde}(1969)}]{heyde1969}%
  \BibitemOpen
  \bibfield  {author} {\bibinfo {author} {\bibfnamefont {C.}~\bibnamefont
  {Heyde}}\ and\ \bibinfo {author} {\bibfnamefont {E.}~\bibnamefont {Heyde}},\
  }\href {\doibase 10.1016/S0022-5193(69)80022-6} {\bibfield  {journal}
  {\bibinfo  {journal} {J. Theor. Biol.}\ }\textbf {\bibinfo {volume} {25}},\
  \bibinfo {pages} {159} (\bibinfo {year} {1969})}\BibitemShut {NoStop}%
\bibitem [{\citenamefont {{van Kampen}}(2007)}]{vanKampen2007}%
  \BibitemOpen
  \bibfield  {author} {\bibinfo {author} {\bibfnamefont {N.~G.}\ \bibnamefont
  {{van Kampen}}},\ }\href@noop {} {\emph {\bibinfo {title} {Stochastic
  Processes in Physics and Chemistry}}},\ \bibinfo {edition} {3rd}\ ed.\
  (\bibinfo  {publisher} {Elsevier},\ \bibinfo {address} {New York},\ \bibinfo
  {year} {2007})\BibitemShut {NoStop}%
\bibitem [{\citenamefont {Doob}(1945)}]{doob1945}%
  \BibitemOpen
  \bibfield  {author} {\bibinfo {author} {\bibfnamefont {J.~L.}\ \bibnamefont
  {Doob}},\ }\href {http://www.jstor.org/stable/1990339} {\bibfield  {journal}
  {\bibinfo  {journal} {Trans. Amer. Math. Soc.}\ }\textbf {\bibinfo {volume}
  {58}},\ \bibinfo {pages} {455} (\bibinfo {year} {1945})}\BibitemShut
  {NoStop}%
\bibitem [{\citenamefont {Gillespie}(1976)}]{gillespie1976}%
  \BibitemOpen
  \bibfield  {author} {\bibinfo {author} {\bibfnamefont {D.~T.}\ \bibnamefont
  {Gillespie}},\ }\href {\doibase 10.1016/0021-9991(76)90041-3} {\bibfield
  {journal} {\bibinfo  {journal} {J. Comput. Phys.}\ }\textbf {\bibinfo
  {volume} {22}},\ \bibinfo {pages} {403} (\bibinfo {year} {1976})}\BibitemShut
  {NoStop}%
\bibitem [{\citenamefont {Gillespie}(1977)}]{gillespie1977}%
  \BibitemOpen
  \bibfield  {author} {\bibinfo {author} {\bibfnamefont {D.~T.}\ \bibnamefont
  {Gillespie}},\ }\href {\doibase 10.1021/j100540a008} {\bibfield  {journal}
  {\bibinfo  {journal} {J. Phys. Chem.}\ }\textbf {\bibinfo {volume} {81}},\
  \bibinfo {pages} {2340} (\bibinfo {year} {1977})}\BibitemShut {NoStop}%
\bibitem [{\citenamefont {Daley}\ and\ \citenamefont
  {Vere-Jones}(2003)}]{daley2003}%
  \BibitemOpen
  \bibfield  {author} {\bibinfo {author} {\bibfnamefont {D.~J.}\ \bibnamefont
  {Daley}}\ and\ \bibinfo {author} {\bibfnamefont {D.}~\bibnamefont
  {Vere-Jones}},\ }\href@noop {} {\emph {\bibinfo {title} {An Introduction to
  the Theory of Point Processes}}},\ \bibinfo {edition} {2nd}\ ed.,\
  Vol.~\bibinfo {volume} {1}\ (\bibinfo  {publisher} {Springer},\ \bibinfo
  {address} {New York},\ \bibinfo {year} {2003})\BibitemShut {NoStop}%
\bibitem [{\citenamefont {Bhabha}(1950)}]{bhabha1950}%
  \BibitemOpen
  \bibfield  {author} {\bibinfo {author} {\bibfnamefont {H.~J.}\ \bibnamefont
  {Bhabha}},\ }\href {\doibase 10.1098/rspa.1950.0102} {\bibfield  {journal}
  {\bibinfo  {journal} {Proc. Roy. Soc. A}\ }\textbf {\bibinfo {volume}
  {202}},\ \bibinfo {pages} {300} (\bibinfo {year} {1950})}\BibitemShut
  {NoStop}%
\bibitem [{\citenamefont {Ramakrishnan}(1950)}]{ramakrishnan1950}%
  \BibitemOpen
  \bibfield  {author} {\bibinfo {author} {\bibfnamefont {A.}~\bibnamefont
  {Ramakrishnan}},\ }\href {\doibase 10.1017/S0305004100026153} {\bibfield
  {journal} {\bibinfo  {journal} {Math. Proc. Cambridge}\ }\textbf {\bibinfo
  {volume} {46}},\ \bibinfo {pages} {595} (\bibinfo {year} {1950})}\BibitemShut
  {NoStop}%
\bibitem [{SI()}]{SI}%
  \BibitemOpen
  \href@noop {} {\enquote {\bibinfo {title} {See supplemental material at [url
  will be inserted by publisher] for detailed steps of analytical
  derivation.}}\ }\BibitemShut {NoStop}%
\bibitem [{\citenamefont {Lu}\ \emph {et~al.}(1998)\citenamefont {Lu},
  \citenamefont {Xun},\ and\ \citenamefont {Xie}}]{lu1998}%
  \BibitemOpen
  \bibfield  {author} {\bibinfo {author} {\bibfnamefont {H.~P.}\ \bibnamefont
  {Lu}}, \bibinfo {author} {\bibfnamefont {L.}~\bibnamefont {Xun}}, \ and\
  \bibinfo {author} {\bibfnamefont {X.~S.}\ \bibnamefont {Xie}},\ }\href
  {\doibase 10.1126/science.282.5395.1877} {\bibfield  {journal} {\bibinfo
  {journal} {Science}\ }\textbf {\bibinfo {volume} {282}},\ \bibinfo {pages}
  {1877} (\bibinfo {year} {1998})}\BibitemShut {NoStop}%
\bibitem [{\citenamefont {Bartholomay}(1962)}]{bartholomay1962}%
  \BibitemOpen
  \bibfield  {author} {\bibinfo {author} {\bibfnamefont {A.~F.}\ \bibnamefont
  {Bartholomay}},\ }\href {\doibase 10.1021/bi00908a005} {\bibfield  {journal}
  {\bibinfo  {journal} {Biochemistry}\ }\textbf {\bibinfo {volume} {1}},\
  \bibinfo {pages} {223} (\bibinfo {year} {1962})}\BibitemShut {NoStop}%
\end{thebibliography}%
\end{document}